\documentclass[12pt,a4paper,twoside]{article}

\usepackage{graphicx}
\usepackage{color}
\usepackage{lscape}
\usepackage{hyperref}
\usepackage{epstopdf}
\usepackage{lscape}
\usepackage{cite}
\usepackage{amsmath}
\usepackage{amsfonts}
\usepackage{amssymb}

\usepackage{hyperref}
\usepackage{amsmath}
\usepackage{setspace}
\begin{document}
\begin{center}
\textbf{{\Large Quantum Information Entropies for the $\ell$-state P\"oschl-Teller-type potential}}\\
\vspace{5mm}

\end{center}
\begin{center}
\textbf{Yahya, W. A.$^\dagger$ $^\ddagger$\footnote{wazzy4real@yahoo.com} , Oyewumi, K. J.$^\dagger$\footnote{mjpysics@yahoo.com} , Sen, K. D$^\star$ \footnote{sensc@uohyd.ernet.in} }.\\
\vspace{4mm}
 $^\dagger$ { Theoretical Physics Section, Department of Physics, University of Ilorin, Nigeria. \\}

$^\ddagger$ { Department of Physics and Material Science, Kwara
State University, Malete, Nigeria. \\}

$^\star$ { School of Chemistry, University of Hyderabad, Hyderabad 500046, India \\} 
\end{center}
\vspace{4mm}
\begin{abstract}
In this study, we obtained the position-momentum uncertainties and some uncertainty relations for the P\"oschl-Teller-type potential for any $\ell$. The radial expectation values of $r^{-2}$, $r^{2}$ and $p^{2}$ are obtained from which the Heisenberg Uncertainty principle holds for the potential model under consideration. The Fisher information is then obtained and it is observed that the Fisher-information-based uncertainty relation and the Cramer-Rao inequality hold for this even power potential. Some numerical and graphical results are displayed. 
\end{abstract}

\vspace{4mm}3
{\bf PACs No.}: 03.65.Fd, 03.65.Ge, 03.65.Ca, 03.65-W.

\section{Introduction}
The quantum mechanical uncertainty principle, first formulated in terms of the standard deviations of the position and momentum probability densities which characterize the quantum-mechanical states of one-dimensional single particle systems, is fundamental to understanding the electronic structure and properties of atoms are molecules \cite{GoE09}. In particular, we have the Heisenberg uncertainty principle \cite{Hei27} for the product of the uncertainties in position and momentum, expressed in terms of Planck's constant \cite{PaS071, PaS12}. For a one-dimensional system defined over $-\infty \leq x \leq \infty$, it is given by the product of the corresponding uncertainties
\begin{equation}
(\Delta x)=\sqrt{<x^{2}>-<x>^{2}} \; \; \mathrm{and} \; \; (\Delta p)=\sqrt{<p^{2}>-<p>^{2}},
\label{u1}
\end{equation}
according to
\begin{equation}
(\Delta x) (\Delta p)\geq\frac{\hbar}{2}.
\label{u2}
\end{equation}

The Fisher information entropy is defined in position space as \cite{DeE05,KaS10}
\begin{equation}
I[\rho]=\int \frac{\left[ \nabla \rho(\textbf{r})\right] ^{2}}{\rho (\textbf{r})}d\textbf{r},
\label{u3}
\end{equation}
and the corresponding momentum space measure is given by \cite{KaS10}
\begin{equation}
I[\gamma]=\int \frac{\left[ \nabla \gamma(\textbf{p})\right] ^{2}}{\gamma (\textbf{p})}d\textbf{p},
\label{u4}
\end{equation}
where $\rho$ and $\gamma$ are the probability densities in the position and momentum spaces, respectively. The importance of Fisher information as a measure of the information in a distribution is well known. It has many implications in estimation theory, as exemplified by the Cramer-Rao bound which is a fundamental limit on the variance of an estimator \cite{BeV09}. It is a useful tool for characterizing complex signals or systems with applications in geophysics, biology and so on, see refs. \cite{BeV09, Fri04} for details. 

A sharp (smooth) and strongly localized (well spread out) probability density gives rise to a larger(smaller) value of the Fisher information in the position space \cite{KaS10}. The Fisher information entropy in the position space measures the narrowness and the oscillation nature of the probability distribution \cite{MoS08}.

The individual Fisher measures are bounded through the Cramer-Rao inequality \cite{Rao65} according to 
\begin{equation}
I[\rho] \geq \frac{1}{\sigma^{2}[\rho]}, \; \; I[\gamma] \geq \frac{1}{\sigma^{2}[\gamma]},
\label{u5}
\end{equation}
where $\sigma^{2}[\rho]$ and $\sigma^{2}[\gamma]$ are, respectively, the standard deviation in the position and momentum spaces \cite{PaS071,RoE05}. Different properties of some quantum mechanical potentials have been studied using the Heisenberg uncertainty principle \cite{PaS072}. Also, Dehesa et al. (2006) studied the information-theoretic measures for Morse and P\"oschl-Teller potentials. It was observed that the ground state of these
potentials saturates all the uncertainty relations in an appropriate limit of the parameter \cite{DeE06}.

In this study, we shall study the Heisenberg uncertainty principle and the quantum information entropies for the P\"oschl-Teller-type potential for any $\ell$ . As earlier stated, for a particle moving non-relativistically in a central potential $V(r)$, the following uncertainty relation holds $(\langle x \rangle = \langle p \rangle = 0)$ \cite{QiD06, DoM89, Nie79, Thi79, Wol74, Pop99, GrE04, OyE04}:
\begin{equation}
\langle r^{2} \rangle_{n \ell} \langle p^{2} \rangle_{n \ell} \geq \frac{9}{4} \hbar^{2}
\label{u6}.
\end{equation}
The inequality above becomes equality for the ground state of the harmonic oscillator (HO). The extent to which the above inequality is saturated depends on the potential shape and the state $(n, \ell)$ of interest. For harmonic oscillator, the inequality in (\ref{u6}) attains its minimum value $2.25$ for $n = \ell = 0$ and $\hbar = 1$. Only a few of these central potentials can be solved analytically for the above inequality for any bound state $n, \ell$.

The P\"oschl-Teller-type potential considered in this study is \cite{ChE13}:
\begin{equation}
V(r)=\frac{\hbar^{2}\alpha^{2}\lambda(\lambda+1)}{2M}\tanh^{2}(\alpha r), \: \: (-\infty<r<\infty),
\label{u7}
\end{equation}
where $\mu$ is the mass of the particle, and $\lambda$ denotes the potential depth and $\alpha$ is related to the range of the potential. This potential is a non-homogeneous potential, as described by Sen and Katriel (2006) \cite{SeK06}. It belongs to a class of even-power series potentials, this class of potentials behave like a harmonic oscillator potential (near the origin), often termed `oscillator-like' potentials \cite{LiG97, GrE04, OyE04}.

The study is organized as follows: In Section \ref{sec2}, we obtain the expectation values of $r^{-2}$, $r^{2}$ and $p^{2}$ from which the Heisenberg uncertainty product is obtained for the P\"oschl-Teller-type potential for any $\ell$ . Section \ref{sec3} contains the Fisher information measure and the Cramer-Rao product of the P\"oschl-Teller-type potential for any $\ell$. The conclusion is given in Section \ref{sec4}.

\section{Position-momentum uncertainty relations}
\label{sec2}
The energy eigenvalues for the $\ell$-state P\"oschl-Teller-type potential is obtained as \cite{YaO14}
\begin{equation}
E_{n \ell}=\frac{\hbar^{2}}{2\mu}\left[ 4d_{0} \Lambda-4n^{2}-4n+8n\gamma+4\gamma-8n\zeta-4\zeta-\frac{3}{2}-\Lambda+8\gamma \zeta\right],
\label{u8}
\end{equation}
where
\begin{equation}
\Lambda=\alpha^{2}\ell(\ell+1),\; \; \beta=\alpha^{2}\lambda(\lambda+1), \; s = \sinh^{2}{(\alpha r)},  \;  \gamma=\sqrt{\frac{1}{16}+\frac{\beta}{4}}, \; \; \zeta=\sqrt{\frac{1}{16}+\frac{\Lambda}{4}}.
\label{u9}
\end{equation}
Also, $n=0,1,2,...,[\lambda]$, where $[\lambda]$ denotes the largest integer inferior to $\lambda$. The wave function is obtained as \cite{YaO14}
\begin{equation}
R_{n\ell}=s^{\frac{1}{4}+\zeta}(1+s)^{\frac{1}{4}-\gamma}P_{n}^{(2\zeta,-2\gamma)}(1+2s).
\label{u10}
\end{equation}

\begin{figure}[!h]
\caption{Graph of $<r^{-2}> _{n \ell}$ against $\lambda$ for some $n$ and $\ell$}
\centering
\includegraphics[scale=1]{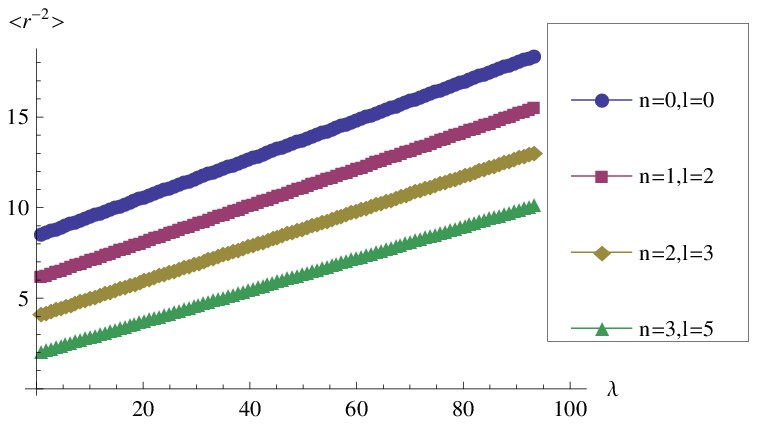}
\label{fig1}
\end{figure}

The expectation value $<r^{-2}>$ of the P\"oschl-Teller-type potential is calculated explicitly, by using the Hellmann-Feynman theorem theorem (HFT)\cite{GrE04, OyE04, Pop98, Hel37, Fey39, EpE62, MaM85, Wal05, OyE08, FlE11, OyS12}. The HFT states that a non-degenerate eigenvalue $E(q)$ of a parameter-dependent Hermitian operator $H(q)$, the associated eigenvector $\Psi(q)$, changes with respect to the parameter $q$ according to the formula \cite{FlE11, OyS12}:
\begin{equation}
\frac{\partial E_{n,\ell}(q)}{\partial q}=\left\langle \Psi_{n,\ell}(q) | \frac{\partial H(q)}{\partial q} | \Psi_{n,\ell}(q) \right\rangle .
\label{u11}
\end{equation}
The Hamiltonian of the system is
\begin{equation}
H=\frac{p^{2}}{2\mu}+\frac{\hbar^{2}\alpha^{2}}{2\mu}\frac{\ell(\ell+1)}{r^{2}}+\frac{\hbar^{2}\alpha^{2}}{2\mu}\frac{\lambda(\lambda+1)}{r^{2}}\tanh^{2}\alpha r
\label{u12}
\end{equation}
and by putting $q = \ell$, we obtain
\begin{equation}
\left\langle r^{-2} \right\rangle _{n \ell}=-\frac{2}{3}-\frac{n}{\sqrt{\zeta}}-\frac{1}{2 \sqrt{\zeta}}+\frac{\gamma}{\sqrt{\zeta}}. 
\label{u13}
\end{equation}
The graph of $\left\langle r^{-2} \right\rangle _{n \ell}$ against $\lambda$ for some values of $n$ and $\ell$  is plotted in Figure \ref{fig1}. We observe that $\left\langle r^{-2} \right\rangle _{n \ell}$ increases linearly with $\lambda$. The values are also shown in Table \ref{tab1} for some $n$ and $\ell$. It is observed from Table \ref{tab1} that $\left\langle r^{-2} \right\rangle _{n \ell}$ decreases with increasing $n$ when $\ell$ is fixed. 

In order to obtain the uncertainty product $(\Delta r)^{2}_{n \ell} (\Delta p)^{2}_{n \ell}$, we need to calculate the following expectation values: $ \langle r \rangle_{n \ell}, \langle p \rangle_{n \ell}, \langle r^2 \rangle_{n \ell} ~\mbox{and}~ \langle p^2 \rangle_{n \ell}$. As expected for a particle moving in the symmetric potential well, both $\langle r \rangle_{n \ell}$ and $\langle p \rangle_{n \ell}$ are zero \cite{ChE13}. $\langle r^2 \rangle_{n \ell}$ is obtained as
\begin{eqnarray}
\left \langle r^{2}\right\rangle _{n \ell} & = & 4 \pi \int_{-\infty}^{\infty}r^{2}R_{n \ell}^{\ast}(r)r^{2}R_{n \ell}(r)dr \nonumber \\
& = & 4 \pi N^{2}\int_{0}^{1}\frac{(\sinh^{-1}\sqrt{s})^{4} }{2 \alpha^{5}}s^{2\zeta}(1+s)^{-2\gamma}\left[P_{n}^{(2\zeta,-2\gamma)}(1+2s) \right] ^{2}ds.
\label{u14}
\end{eqnarray}
From the Hamiltonian of the system, we obtain
\begin{eqnarray}
\langle p^{2}\rangle_{n \ell} & = & 4 \pi \int_{-\infty}^{\infty}r^{2}R_{n \ell}^{\ast}(r)\left[2\mu H-\frac{\hbar^{2} \Lambda}{r^{2}}-\hbar^{2} \beta \tanh^{2}\alpha r \right]R_{n \ell}(r)dr \nonumber \\
& = & 4 \pi\left[2\mu E_{n,\ell}- \hbar^{2} \Lambda \int_{-\infty}^{\infty} \! \! \! R_{n \ell}^{2}(r)dr-\hbar^{2} \beta \int_{-\infty}^{\infty} \! \! \! r^{2}\tanh^{2}\alpha r R_{n \ell}^{2}(r)dr \right],
\label{u15}
\end{eqnarray}
so that
\begin{eqnarray}
\langle p^{2}\rangle_{n \ell} & = & 4 \pi \left\lbrace 2\mu E_{n,\ell}- \frac{\hbar^{2} \Lambda N^{2}}{2 \alpha}\int_{0}^{1} \! \! \! s^{2 \zeta}(1+s)^{-2 \gamma}\left[ P_{n}^{(2 \zeta,-2 \gamma)}(1+2s)\right]^{2}ds \right. \nonumber \\
& - & \left.  \! \! \! \! \frac{\hbar^{2} \beta N^{2}}{2\alpha^{3}}\int_{0}^{1} \! \! \! s^{1+2\zeta}(1+s)^{-1-2\gamma}\left(\sinh^{-1}\sqrt{s} \right) ^{2} \left[ P_{n}^{(2 \zeta,-2 \gamma)}(1+2s)\right]^{2}ds \right\rbrace.
\label{u16}
\end{eqnarray}

A state is defined to be squeezed if $(\Delta r)^{2}<0.5$. Our results from Tables \ref{tab2}-\ref{tab6} show a squeezed phenomenon in the position $r$ for the ground state when $\lambda\geq2.5$. For the first excited state $(n=1)$, we have squeezed phenomena from $\lambda = 2.2$ to $\lambda=5.3$, also at $\lambda\geq 11.4$. We also have squeezed phenomena at various points for the states $(n = 2, \ell=0), \,(n = 2, \ell=1),\,(n = 3, \ell=2)$. It is observed from Tables \ref{tab2} - \ref{tab6} that the lower bounds for the single particle in any central potential is obtained as \cite{Fri04,RoE05}
\begin{equation}
(\Delta r)^{2}_{n \ell} (\Delta p)^{2}_{n \ell} = \left\langle r^{2} \right\rangle _{n \ell}  \left\langle p^{2} \right\rangle _{n \ell}  \geq \left(\ell+\frac{3}{2} \right) ^{2}
\label{u17}
\end{equation}
holds for this potential.  The least value for $\ell = 0$ with various values of $n$ obtained is $3.306209$ which is greater than the expected minimum value $(2.250)$ as shown in Tables \ref{tab2} - \ref{tab4}. For various values of $n~ \mbox{and} ~\ell$, the Heisenberg uncertainty principle  holds for the various values of $\lambda$ considered as it can be seen from Tables \ref{tab5} - \ref{tab9}, since the least value obtained is greater than the minimum value of equation (\ref{u17}).

\section{Fisher Information}
\label{sec3}
The Fisher information of the P\"oschl-Teller-type potential for any $\ell$. Noting that the Fisher information for the expectation values of $r$ and $p$ can be obtained by using the following relations: \cite{PaS071}
\begin{equation}
I[\rho]=4 \left\langle p^{2} \right\rangle _{n \ell}  - 2(2\ell+1)|m| \left\langle  r^{-2} \right\rangle _{n \ell} 
\label{u18}
\end{equation}
and
\begin{equation}
I[\gamma]=4 \left\langle r^{2} \right\rangle _{n \ell}  - 2(2\ell+1)|m| \left\langle  p^{-2} \right\rangle _{n \ell}.
\label{u19}
\end{equation}
The $\langle r^{2} \rangle$ and $\langle p^{2} \rangle$ have been obtained in equations (\ref{u14}) and (\ref{u16}), respectively.  On substituting equations (\ref{u14}) and (\ref{u16}) into equations (\ref{u18}) and (\ref{u19}), the Fisher information in the position and momentum spaces are obtained. In this case, the magnetic quantum number $(m)$ is zero.  According to Dehesa et al. (2007), the product of both equations above can not be less than $36.00$ \cite{DeE07}:
\begin{equation}
I[\rho] \times I[\gamma] \geq 36.00
\label{u20}.
\end{equation}
 The values obtained are shown in Tables \ref{tab10} - \ref{tab14}. The Fisher-information-based uncertainty relation holds for this potential model, since the minimum value of $I[\rho]\times I[\gamma]$ from the Tables is $52.89935$ which is greater than the expected minimum value of $36$ (from equation (\ref{u20})). Similarly,  the Cramer-Rao inequality  given by Dehesa et al. (2007) \cite{DeE07}
\begin{equation}
I[\rho] \times V[\rho] \geq D^2
\label{u21}
\end{equation}
 is satisfied for this potential as shown in Table \ref{tab15}. In this case, $D$ is $3$, so that the minimum value is $9.00$.

\section{Conclusion}
\label{sec4}
We have studied some entropic-uncertainty relations for the P\"oschl-Teller-type potential for any $\ell$. We obtained the expectation values of $r^{-2}$, $r^{2}$ and $p^{2}$. The Heisenberg Uncertainty principle for any $\ell$ is satisfied for this potential model as seen in the numerical results. 
The Fisher information is then obtained by making use of the expectation values of $r^2$ and $p^2$, coupled with the fact that the magnetic quantum number $m$ is zero in the case considered. The validity of the Heisenberg, Fisher-information-based uncertainty relations and Cramer-Rao products for the P\"{o}schl-Teller type has been investigated. We established that the condition of the Fisher-Information-based uncertainty product is satisfied  for the P\"{o}schl-Teller type potential model. Also, the Cramer-Rao inequality holds for this potential model.

\begin{table}[!h]
\centering
\caption{{ Numerical results of $<r^{-2}>_{n \ell}$ for the P\"oschl-Teller-type potential with $\lambda=100$, $\hbar = 2 \mu = 1$, $\alpha=0.1$ for some $n$ and $\ell$}}
\begin{tabular}{ccc}
\hline \hline
$n$ & $\ell$ & $<r^{-2}>_{n \ell}$ \\ [1ex]
\hline \hline
0 & 0 & 8.39564 \\ [1ex]

1 & 0 & 6.39564 \\ [1ex]

2 & 0 & 4.39564 \\ [1ex]

3 & 0 & 2.39564 \\ [1ex]

1 & 2 & 6.02588 \\ [1ex]

2 & 2 & 4.13059 \\ [1ex]

3 & 2 & 2.23530 \\ [1ex]

4 & 2 & 0.34002 \\ [1ex]

1 & 5 & 5.13217 \\ [1ex]

2 & 5 & 3.48998 \\ [1ex]

3 & 5 & 1.84778 \\ [1ex]

4 & 5 & 0.20559 \\ [1ex]
\hline \hline
\end{tabular} 
\label{tab1}
\end{table}

\begin{table}[!h]
\caption{Numerical results for the uncertainty relation $(\Delta r)^{2}_{n \ell} (\Delta p)^{2}_{n \ell}$ with $\hbar = 2 \mu = 1$, $\alpha=1$, $n = 0$ and $\ell=0$}
\begin{tabular}{ccccc}
\hline \hline 
$\lambda$ & $<r^{2}>_{n \ell}$ & $<p^{2}>_{n \ell}$ & $(\Delta r)^{2}_{n \ell} (\Delta p)^{2}_{n \ell}$ & min $((\Delta r)^{2}_{n \ell} (\Delta p)^{2}_{n \ell})$ \\ 
\hline \hline
0.5 & 0.551859 & 5.99104 & 3.306209 & 2.25 \\ [1ex] 

1.5 & 0.526554 & 42.5751 & 22.41728 & 2.25 \\ [1ex]

2.5 & 0.498741 & 78.5373 & 39.16977 & 2.25 \\ [1ex]

3.5 & 0.468876 & 113.996 & 53.44999 & 2.25 \\ [1ex]

4.5 & 0.437563 & 149.079 & 65.23145 & 2.25 \\ [1ex]

5.5 & 0.405598 & 183.915 & 74.59556 & 2.25 \\ [1ex]

6.5 & 0.373856 & 218.619 & 81.73202 & 2.25 \\ [1ex]

7.5 & 0.343174 & 253.286 & 86.92117 & 2.25 \\ [1ex]

8.5 & 0.314248 & 287.981 & 90.49745 & 2.25 \\ [1ex]

9.5 & 0.287563 & 322.742 & 92.80866 & 2.25 \\ [1ex]
\hline \hline 
\end{tabular} 
\label{tab2}
\end{table}

\begin{table}[!h]
\caption{Numerical results for the uncertainty relation $(\Delta r)^{2}_{n \ell} (\Delta p)^{2}_{n \ell}$ with $\hbar = 2 \mu = 1$, $\alpha=1$, $n = 1$ and $\ell=0$}
\begin{tabular}{ccccc}
\hline \hline
$\lambda$ & $<r^{2}>_{n \ell}$ & $<p^{2}>_{n \ell}$ & $(\Delta r)^{2}_{n \ell} (\Delta p)^{2}_{n \ell}$ & min $((\Delta r)^{2}_{n \ell} (\Delta p)^{2}_{n \ell})$ \\ 
\hline \hline 
1.5 & 0.548912 & 17.3952 & 9.548434 & 2.25 \\ [1ex]

2.5 & 0.461496 & 103.858 & 47.93005 & 2.25 \\ [1ex]

3.5 & 0.302590 & 190.969 & 57.78531 & 2.25 \\ [1ex]

4.5 & 0.334296 & 276.538 & 92.44555 & 2.25 \\ [1ex]

5.5 & 0.516568 & 357.880 & 184.8694 & 2.25 \\ [1ex]

6.5 & 0.569420 & 439.477 & 250.2470 & 2.25 \\ [1ex]

7.5 & 0.572668 & 521.226 & 298.4895 & 2.25 \\ [1ex]

8.5 & 0.560332 & 602.821 & 337.7799 & 2.25 \\ [1ex]

9.5 & 0.541791 & 684.221 & 370.7048 & 2.25 \\ [1ex]
\hline \hline
\end{tabular} 
\label{tab3}
\end{table}

\begin{table}[!h]
\caption{Numerical results for the uncertainty relation $(\Delta r)^{2}_{n \ell} (\Delta p)^{2}_{n \ell}$ with $\hbar = 2 \mu = 1$, $\alpha=1$, $n = 2$ and $\ell=0$}
\begin{tabular}{ccccc}
\hline \hline
$\lambda$ & $<r^{2}>_{n \ell}$ & $<p^{2}>_{n \ell}$ & $(\Delta r)^{2}_{n \ell}(\Delta p)^{2}_{n \ell}$ & min $((\Delta r)^{2}_{n \ell} (\Delta p)^{2}_{n \ell})$ \\ 
\hline \hline 
2.5 & 0.550407 & 28.0161 & 15.4203 & 2.25 \\ [1ex]

3.5 & 0.362748 & 165.251 & 59.9445 & 2.25 \\ [1ex]
 
4.5 & 0.411899 & 300.560 & 123.800 & 2.25 \\ [1ex]
 
5.5 & 0.561492 & 432.187 & 242.670 & 2.25 \\ [1ex]
 
6.5 & 0.540564 & 565.700 & 305.797 & 2.25 \\ [1ex]
 
7.5 & 0.476402 & 700.333 & 333.640 & 2.25 \\ [1ex]
 
8.5 & 0.396338 & 836.261 & 331.442 & 2.25 \\ [1ex]
 
9.5 & 0.369105 & 970.598 & 358.253 & 2.25 \\ [1ex]
\hline \hline 
\end{tabular} 
\label{tab4}
\end{table}

\begin{table}[!h]
\caption{Numerical results for the uncertainty relation $(\Delta r)^{2}_{n \ell} (\Delta p)^{2}_{n \ell}$ with $\hbar = 2 \mu = 1$, $\alpha=1$, $n = 2$ and  $\ell=1$}
\begin{tabular}{ccccc}
\hline \hline 
$\lambda$ & $<r^{2}>_{n \ell}$ & $<p^{2}>_{n \ell}$ & $(\Delta r)^{2}_{n \ell}(\Delta p)^{2}_{n \ell}$ & min $((\Delta r)^{2}_{n \ell} (\Delta p)^{2}_{n \ell})$ \\ 
\hline \hline
3.5 & 0.578148 & 117.507 & 67.93561 & 6.25 \\ [1ex]

4.5 & 0.484400 & 277.361 & 134.3537 & 6.25 \\ [1ex]

5.5 & 0.317118 & 437.433 & 138.7170 & 6.25 \\ [1ex]

6.5 & 0.418443 & 595.168 & 249.0439 & 6.25 \\ [1ex]

7.5 & 0.534586 & 751.435 & 401.7066 & 6.25 \\ [1ex]

8.5 & 0.538793 & 908.354 & 489.4148 & 6.25 \\ [1ex]

9.5 & 0.498933 & 1066.09 & 531.9075 & 6.25 \\ [1ex]

10.5 & 0.437253 & 1225.02 & 535.6437 & 6.25 \\ [1ex]
\hline \hline
\end{tabular} 
\label{tab5}
\end{table}

\begin{table}[!h]
\caption{Numerical results for the uncertainty relation $(\Delta r)^{2}_{n \ell} (\Delta p)^{2}_{n \ell}$ with $\hbar = 2 \mu = 1$, $\alpha=1$, $n = 3$ and $\ell=2$}
\begin{tabular}{ccccc}
\hline \hline 
$\lambda$ & $<r^{2}>_{n \ell}$ & $<p^{2}>_{n \ell}$ & $(\Delta r)^{2}_{n \ell}(\Delta p)^{2}_{n \ell}$ & min $((\Delta r)^{2}_{n \ell} (\Delta p)^{2}_{n \ell})$ \\ 
\hline \hline
4.5 & 0.639235 & 61.0028 & 38.9951 & 12.25 \\ [1ex] 

5.5 & 0.596992 & 294.683 & 175.923 & 12.25 \\ [1ex]

6.5 & 0.503234 & 527.632 & 265.522 & 12.25 \\ [1ex]

7.5 & 0.340425 & 759.800 & 258.655 & 12.25 \\ [1ex]

8.5 & 0.449519 & 990.761 & 445.366 & 12.25 \\ [1ex]

9.5 & 0.528862 & 1221.84 & 646.185 & 12.25 \\ [1ex]

10.5 & 0.503222 & 1453.55 & 731.458 & 12.25 \\ [1ex]
\hline \hline
\end{tabular} 
\label{tab6}
\end{table}

\begin{table}[!h]
\caption{The values of $(\Delta r)_{n \ell}$ for some $\lambda$ with $\hbar = 2 \mu = 1$ and $\alpha=1$}
\begin{tabular}{cccccc}
\hline \hline 
$n$ & $\ell$ & $\lambda=20$ & $\lambda=50$ & $\lambda=100$ & $\lambda=200$ \\ [1ex]
\hline 
3 & 0 & 0.750849 & 0.508588 & 0.347300 & 0.241542 \\ [1ex]

4 & 0 & 0.663283 & 0.583597 & 0.394231 & 0.272851 \\ [1ex]

5 & 0 & 0.706252 & 0.654627 & 0.437750 & 0.301451 \\ [1ex]

6 & 1 & 0.678838 & 0.739476 & 0.497916 & 0.340207 \\ [1ex]

7 & 1 & 0.707238 & 0.751129 & 0.536695 & 0.364714 \\ [1ex]

8 & 1 & 0.716054 & 0.690046 & 0.574314 & 0.388097 \\ [1ex]
\hline \hline
\end{tabular} 
\label{tab7}
\end{table}

\begin{table}[!h]
\caption{The values of $(\Delta p)_{n \ell}$ for some $\lambda$ with $\hbar = 2 \mu = 1$ and $\alpha=1$}
\begin{tabular}{cccccc}
\hline \hline
$n$ & $\ell$ & $\lambda=20$ & $\lambda=50$ & $\lambda=100$ & $\lambda=200$ \\ [1ex]
\hline
3 & 0 & 54.6780 & 90.8740 & 130.833 & 186.646 \\ [1ex]

4 & 0 & 60.1654 & 101.147 & 146.469 & 209.523 \\ [1ex]

5 & 0 & 64.1146 & 110.033 & 160.284 & 229.923 \\ [1ex]

6 & 1 & 68.4516 & 121.564 & 178.512 & 257.144 \\ [1ex]

7 & 1 & 70.2057 & 128.531 & 189.370 & 273.577 \\ [1ex]

8 & 1 & 71.2527 & 135.263 & 199.394 & 288.908 \\ [1ex]
\hline \hline
\end{tabular} 
\label{tab8}
\end{table}

\begin{table}[!h]
\caption{The values of $(\Delta r \Delta p)_{n \ell}$ for some $\lambda$ with $\hbar = 2 \mu = 1$ and $\alpha=1$}
\begin{tabular}{ccccccc}
\hline \hline
$n$ & $\ell$ & $\lambda=20$ & $\lambda=50$ & $\lambda=100$ & $\lambda=200$ & min$(\Delta r \Delta p)_{n \ell}$ \\ [1ex]
\hline 
3 & 0 & 41.0549 & 46.2170 & 45.4383 & 45.0828 & 1.50 \\ [1ex]

4 & 0 & 39.9067 & 59.0290 & 57.7426 & 57.1686 & 1.50 \\ [1ex]

5 & 0 & 45.2811 & 72.0306 & 70.1643 & 69.3105 & 1.50 \\ [1ex]

6 & 1 & 46.4675 & 89.8937 & 88.8840 & 87.4822 & 2.50 \\ [1ex]

7 & 1 & 49.6521 & 96.5434 & 101.634 & 99.7770 & 2.50 \\ [1ex]

8 & 1 & 51.0208 & 93.3377 & 114.515 & 112.124 & 2.50 \\ [1ex]
\hline \hline
\end{tabular} 
\label{tab9}
\end{table}

\begin{table}[!h]
\caption{Fisher information measure with $2 \mu = \hbar = \alpha = 1$, $n = 0$ and $\ell=0$}
\begin{tabular}{ccccc}
\hline \hline
$\lambda$ & $I[\rho]$ & $I[\gamma]$ & $I[\rho] I[\gamma]$ & min $(I[\rho] I[\gamma])$ \\ [1ex]
\hline 
0.5 & 23.96416 & 2.207436 & 52.89935 & 36.00 \\ [1ex]

1.5 & 170.3004 & 2.106140 & 358.6765 & 36.00 \\ [1ex]

2.5 & 314.1492 & 1.994964 & 626.7163 & 36.00 \\ [1ex]

3.5 & 455.9840 & 1.875504 & 855.1998 & 36.00 \\ [1ex]

4.5 & 596.3160 & 1.750252 & 1043.703 & 36.00 \\ [1ex]

5.5 & 735.6600 & 1.622392 & 1193.529 & 36.00 \\ [1ex]

6.5 & 874.4760 & 1.495424 & 1307.712 & 36.00 \\ [1ex]

7.5 & 1013.144 & 1.372696 & 1390.739 & 36.00 \\ [1ex]

8.5 & 1151.924 & 1.256992 & 1447.959 & 36.00 \\ [1ex]

9.5 & 1290.968 & 1.150252 & 1484.939 & 36.00 \\ [1ex]
\hline \hline
\end{tabular} 
\label{tab10}
\end{table}

\begin{table}[!h]
\caption{Fisher information measure with $2 \mu = \hbar = \alpha = 1$, $n = 1$ and $\ell=0$}
\begin{tabular}{ccccc}
\hline \hline
$\lambda$ & $I[\rho]$ & $I[\gamma]$ & $I[\rho] I[\gamma]$ & min $(I[\rho] I[\gamma])$ \\ [1ex]
\hline 
1.5 & 69.58080 & 2.195648 & 152.7749 & 36.00 \\ [1ex]

2.5 & 415.4302 & 1.845984 & 766.8808 & 36.00 \\ [1ex]

3.5 & 763.8760 & 1.210360 & 924.5650 & 36.00 \\ [1ex]

4.5 & 1106.152 & 1.337184 & 1479.129 & 36.00 \\ [1ex]

5.5 & 1431.520 & 2.066272 & 2957.910 & 36.00 \\ [1ex]

6.5 & 1757.908 & 2.277680 & 4003.958 & 36.00 \\ [1ex]

7.5 & 2084.904 & 2.290672 & 4775.831 & 36.00 \\ [1ex]

8.5 & 2411.284 & 2.241328 & 5404.478 & 36.00 \\ [1ex]

9.5 & 2736.884 & 2.167164 & 5931.276 & 36.00 \\ [1ex]
\hline \hline
\end{tabular} 
\label{tab11}
\end{table}

\begin{table}[!h]
\caption{Fisher information measure with $2 \mu = \hbar = \alpha = 1$,$n = 2$ and  $\ell=0$}
\begin{tabular}{ccccc}
\hline \hline
$\lambda$ & $I[\rho]$ & $I[\gamma]$ & $I[\rho] I[\gamma]$ & min $(I[\rho] I[\gamma])$ \\ [1ex]
\hline 
2.5 & 112.0644 & 2.201628 & 246.7241 & 36.00 \\ [1ex]

3.5 & 661.0040 & 1.450992 & 959.1115 & 36.00 \\ [1ex]

4.5 & 1202.240 & 1.647596 & 1980.806 & 36.00 \\ [1ex]

5.5 & 1728.748 & 2.245968 & 3882.713 & 36.00 \\ [1ex]

6.5 & 2262.800 & 2.162256 & 4892.753 & 36.00 \\ [1ex]

7.5 & 2801.332 & 1.905608 & 5338.241 & 36.00 \\ [1ex]

8.5 & 3345.044 & 1.585352 & 5303.072 & 36.00 \\ [1ex]

9.5 & 3882.392 & 1.476420 & 5732.041 & 36.00 \\ [1ex]
\hline \hline
\end{tabular} 
\label{tab12}
\end{table}

\begin{table}[!h]
\caption{Fisher information measure with $2 \mu = \hbar = \alpha = 1$, $n = 2$ and $\ell=1$}
\begin{tabular}{ccccc}
\hline \hline
$\lambda$ & $I[\rho]$ & $I[\gamma]$ & $I[\rho] I[\gamma]$ & min $(I[\rho] I[\gamma])$ \\ [1ex]
\hline 
3.5 & 470.0280 & 2.312564 & 1086.970 & 36.00 \\ [1ex]

4.5 & 1109.444 & 1.937600 & 2149.659 & 36.00 \\ [1ex]

5.5 & 1749.732 & 1.268472 & 2219.486 & 36.00 \\ [1ex]

6.5 & 2380.672 & 1.673772 & 3984.702 & 36.00 \\ [1ex]

7.5 & 3005.740 & 2.138344 & 6427.306 & 36.00 \\ [1ex]

8.5 & 3633.416 & 2.155172 & 7830.636 & 36.00 \\ [1ex]

9.5 & 4264.360 & 1.995732 & 8510.520 & 36.00 \\ [1ex]

10.5 & 4900.080 & 1.749012 & 8570.299 & 36.00 \\ [1ex]
\hline \hline
\end{tabular} 
\label{tab13}
\end{table}

\begin{table}[!h]
\caption{Fisher information measure with $2 \mu = \hbar = \alpha = 1$, $n = 3$ and $\ell=2$}
\begin{tabular}{ccccc}
\hline \hline
$\lambda$ & $I[\rho]$ & $I[\gamma]$ & $I[\rho] I[\gamma]$ & min $(I[\rho] I[\gamma])$ \\ [1ex]
\hline 

4.5 & 244.0112 & 2.556940 & 623.9220 & 36.00 \\ [1ex]

5.5 & 1178.732 & 2.387968 & 2814.774 & 36.00 \\ [1ex]

6.5 & 2110.528 & 2.012936 & 4248.479 & 36.00 \\ [1ex]

7.5 & 3039.200 & 1.361700 & 4138.479 & 36.00 \\ [1ex]

8.5 & 3963.044 & 1.798076 & 7125.854 & 36.00 \\ [1ex]

9.5 & 4887.360 & 2.115448 & 10338.96 & 36.00 \\ [1ex]

10.5 & 5814.200 & 2.012888 & 11703.33 & 36.00 \\ [1ex]
\hline \hline
\end{tabular} 
\label{tab14}
\end{table}

\begin{table}[!h]
\caption{The values of the Cramer-Rao product $I[\rho] V[\rho]$ with $2 \mu = \hbar = \alpha = 1$, $n = 3$ and $\ell=2$}
\begin{tabular}{cccccc}
\hline \hline
$\lambda$ & $n=0, \ell=0$ & $n=1, \ell=0$ & $n=2, \ell=1$ & $n=3, \ell=2$ & min $(I[\rho] V[\rho])$ \\ [1ex]
\hline
0.5 & 13.22484 & • & • & • & 9.00 \\ [1ex] 
 
1.5 & 89.66912 & 38.19374 & • & • & 9.00 \\ [1ex]

2.5 & 156.6791 & 191.7202 & • & • & 9.00 \\ [1ex]

3.5 & 213.8000 & 231.1412 & 271.7425 & • & 9.00 \\ [1ex] 

4.5 & 260.9258 & 369.7822 & 537.4147 & 155.9805 & 9.00 \\ [1ex] 

5.5 & 298.3822 & 739.4774 & 554.8715 & 703.6936 & 9.00 \\ [1ex]

6.5 & 326.9281 & 1000.988 & 996.1755 & 1062.089 & 9.00 \\ [1ex]

7.5 & 247.6847 & 1193.958 & 1606.827 & 1034.620 & 9.00 \\ [1ex]

8.5 & 361.9898 & 1351.120 & 1957.659 & 781.4640 & 9.00 \\ [1ex]

9.5 & 371.2346 & 1482.819 & 2127.630 & 2584.739 & 9.00 \\ [1ex]
\hline \hline
\end{tabular} 
\label{tab15}
\end{table}

\begin{thebibliography}{200}
\bibitem{GoE09}	Gonzalez-Ferez, R., Dehesa, J. S., Patil, S. H. and Sen, K. D. (2009): Physica A \textbf{388}, 4919.
\bibitem{Hei27}	Heisenberg, W. (1927): Z. Phys. \textbf{43}, 172.
\bibitem{PaS071} Patil, S. H. and Sen, K. D. (2007): Phys. Lett. A \textbf{370}, 354.
\bibitem{PaS12}	Patil, S. H. and Sen, K. D. (2012): J. Chem. Sci \textbf{124}, 241.
\bibitem{DeE05}	Dehesa, J. S., Lopez-Rosa, S., Olmos, B. and Yanez, R. J. (2005): J. Comput. Appl. Math \textbf{179}, 185.
\bibitem{KaS10} Katriel, J. and Sen, K. D. (2010): J. Comput. Appl. Math. \textbf{233}, 1399.
\bibitem{BeV09} Bercher, J. F. and Vignat, C. (2009): Information Sciences \textbf{179}, 3832-3842. 
\bibitem{Fri04} Frieden, B. R. (2004): \textit{Science from Fisher information}, Cambridge University Press, UK.
\bibitem{MoS08}	Montgomery Jr., H. E. and Sen, K. D. (2008): Phys. Lett. A \textbf{372}, 2271.
\bibitem{Rao65}	Rao, C. R. (1965): \textit{Linear statistical interference and its applications}, Wiley, New York, 1965.
\bibitem{RoE05}	Romera, E., Sanchez-Moreno, P. and Dehesa, J. S. (2005): Chem. Phys. Lett. \textbf{414}, 468.
\bibitem{PaS072} Patil, S. H. and Sen, K. D. (2007): Phys. Lett. A \textbf{362}, 109.
\bibitem{DeE06} Dehesa, J. S., Martinez-Finkelshtein, A. and Sorokin, V. N. (2006): Mol. Phys. {\bf 104} (4), 613.
\bibitem{QiD06} Qiang, W. C. and Dong, S. H. (2006): J. Phys. A: Math. Gen. {\bf 39}, 8663.
\bibitem{DoM89} Dodonov, V. V. and Mank'ko, V. I.(1989): \textit{Generalization of the uncertainty relations in Quantum Mechanics}, Proc. Lebedev Science Institute, {\bf 183}, Nova Science Publication, New York.
\bibitem{Nie79} Nieto, M. M. (1979): Phys. Rev. A {\bf 20}, 700.
\bibitem{Thi79} Thirring, W. (1979): {\it A course in Mathematical Physics. $3$. Quantum Mechanics of Atoms and Molecules}, Springer, New York.
\bibitem{Wol74} Wolsky, A. M. (1974): Am. J. Phys. {\bf 42},  760.
\bibitem{Pop99} Popov, D. (1999): Czech. J. Phys. {\bf 49}(8), 1121.
\bibitem{GrE04} Grypeos, M. E. , Koutroulos, C. G. , Oyewumi, K. J. and Petridou, Th. (2004): J. Phys. A: Math. Gen. \textbf{37}, 7895.
\bibitem{OyE04} Oyewumi, K. J.,  Petridou, Th., Grypeos, M. E. and Koutroulos, C. G. (2004): {\it Approximate Analytic Expressions for the Expectation Values and the Uncertainty Relation for Even-Power- Series Central 
 Potentials using HVT Method.} Proceedings of the 3rd International Workshop on Contemporary Problems 
 in Mathematical Physics, COPROMAPH 3 held on the 1st - 7th Nov. 2003 under the auspices of the 
 International Chair in Mathematical Physics and Applications ICMPA University of Abomey-Calavi, 
 Cotonou, Benin Republic: eds. Profs. J. Govaerts, M.N. Hounkonnou and A. Z. Msezane; 336 - 342. 
\bibitem{ChE13}	Chen, C. Y., You, Y., Lu, F. L. and Dong, S. H. (2013): Phys. Lett. A \textbf{377}, 1070.
\bibitem{SeK06} Sen, K. D. and Katriel, J. (2006): J. Chem. Phys. {\bf 125}, 074117.
\bibitem{YaO14}	Yahya, W. A and Oyewumi, K. J. (2014): Approximate solutions of the $\ell$-state hyperbolic P\"oschl-Teller-type potential. \textit{submitted for publication}.
\bibitem{LiG97} Liolios, T. E. and Grypeos, M. E. (1997): Int. J. Theor. Phys. {\bf 36}, 2051.
\bibitem{Pop98}	Popov, D.(1998): Cze. J. Phys. \textbf{49}(8), 1121.
\bibitem{Hel37}	Hellmann, G. (1937): Einfuhrung in die Quantenchemie (Vienna: Denticke).
\bibitem{Fey39}	Feynman, R. P. (1939): Phys. Rev. \textbf{56}, 340.
\bibitem{EpE62}	Epstein, J. H. and Epstein, S. T. (1962): Am. J. Phys. \textbf{30}, 266.
\bibitem{MaM85}	Marc, G. and McMillan, W. G. (1985): Advances in Chem. Phys. \textbf{58}, 205.
\bibitem{Wal05}	Wallace, D. B.(2005): \textit{An Introduction to Hellmann-Feynman Theory}, M. Sc. Thesis, University of Central Florida, USA.
\bibitem{OyE08} Oyewumi, K. J., Akinpelu, F. O. and Agboola, A. D. (2008): Int. J. Theor. Phys. \textbf{47}(4), 1039. 
\bibitem{FlE11}	Flego, S. P., Plastino, A. and Plastino, A. R. (2011): Annals of Phys. \textbf{326}, 2533.
\bibitem{OyS12}	Oyewumi, K. J. and Sen, K. D. (2012): J. Math. Chem. \textbf{50}, 1039.
\bibitem{DeE07}	Dehesa, J. S., Gonzalez-Ferez, R. and Sanchez-Moreno, P. (2007): J. Phys. A \textbf{40}, 1845.


\end{thebibliography}
\end{document}